\documentclass[prl,aps,superscriptaddress,showpacs,twocolumn,amssymb,amsmath]{revtex4}
\usepackage{graphicx}
\usepackage{theorem}
\usepackage{dcolumn}
\usepackage{bm}
\usepackage{mathrsfs}
\usepackage{setspace}

\begin{document}

\title{Rotation and Angular Momentum Transfer in Bose-Einstein Condensates Induced by Spiral Dark Solitons}

\author{Toshiaki Kanai}
\affiliation{Department of Physics, Osaka City University, 3-3-138 Sugimoto, Sumiyoshi-Ku, Osaka 558-8585, Japan}

\author{Wei Guo \footnote{Email: wguo@magnet.fsu.edu}}
\affiliation{National High Magnetic Field Laboratory, 1800 East Paul Dirac Drive, Tallahassee, FL 32310, USA}
\affiliation{Mechanical Engineering Department, Florida State University, Tallahassee, FL 32310, USA}

\author{Makoto Tsubota}
\affiliation{Department of Physics, Osaka City University, 3-3-138 Sugimoto, Sumiyoshi-Ku, Osaka 558-8585, Japan}
\affiliation{The OCU Advanced Research Institute for Natural Science and Technology (OCARINA), Osaka City University, 3-3-138 Sugimoto, Sumiyoshi-Ku, Osaka 558-8585, Japan}
\date{\today}

\begin{abstract}
It is a common view that rotational motion in a superfluid can exist only in the presence of quantized vortices. However, in our numerical studies on the merging of two concentric Bose-Einstein condensates with axial symmetry in two-dimensional space, we observe the emergence of a spiral dark soliton when one condensate has a non-zero initial angular momentum. This spiral dark soliton enables the transfer of angular momentum between the condensates and allows the merged condensate to rotate even in the absence of quantized vortices. We examine the flow field around the soliton and reveal that its sharp endpoint can induce flow like a vortex point but with a fraction of a quantized circulation. This interesting nontopological ``phase defect'' may generate broad interests since rotational motion is essential in many quantum transport processes.
\end{abstract}
 \maketitle

The hydrodynamics of quantum fluids such as atomic Bose-Einstein condensates (BECs) and superfluid helium are strongly affected by quantum effects \cite{Fetter-2001-JPCM, Donnelly-book, Tsubota-2013-PR}. For instance, it is well-known that in a simply-connected quantum fluid, rotational motion can arise only through the formation of topological defects in the form of quantized vortices, each of which carries a circulation of $\kappa=h/m$, where $h$ is Plancks constant and $m$ is the mass of the particles that form the condensate. In BECs, quantized vortices have been nucleated by a variety of innovative methods, such as direct phase imprint \cite{Matthews-1999-PRL, Leanhardt-2002-PRL}, rotation of the condensate traps \cite{Hodby-2001-PRL, Abo-Shaeer-2001-SCI, Haljan-2001-PRL, Inouye-2001-PRL}, stirring the BECs with laser beams \cite{Madison-2000-PRL} or moving optical obstacles \cite{Sasaki-2010-PRL, Kwon-2016-PRL}, decay of dark solitons \cite{Dutton-2001-SCI, Anderson-2001-PRL}, and merging isolated condensates \cite{Scherer-2007-PRL}. The last method is particularly interesting since it provides a means to test the celebrated Kibble-Zurek mechanism \cite{Weiler-2008-Nature, Carretero-2008-PRA}. This mechanism explains the formation of vortices following a rapid second-order phase transition as due to the merging of isolated superfluid domains with random relative phases \cite{Zurek-1996-PR, Kibble-2007-PT}. In addition, understanding the processes involved in the merging of isolated condensates is also important for matter wave interferometry \cite{Shin-2004-PRL, Hadzibabic-2004-PRL, Yang-2013-PRA, Xiong-2013-PRA}.

So far, many studies on condensate merging have focused on condensates with uniform initial phases. However, the situation is less clear when some condensates contain vortices and have non-uniform phases before merging occurs. Many interesting questions can be raised. For instance, how will the angular momentum be transferred from a rotating condensate to an initially static condensate? Can vortices form due to the velocity gradient near the interface between condensates like that in classical shear flows? Is the angular momentum transfer always associated with vorticity transfer? To provide insights into these questions, we have studied a representative condensate configuration: the merging of a disc Bose-Einstein condensate with a concentric ring condensate in two-dimensional (2D) space, with one of them having a non-zero initial angular momentum induced by a vortex point at the center. We shall report in this Letter the emergence of a spiral dark soliton during the merging process and will show that this soliton can induce rotational motion in the condensate even without quantized vortices.

\begin{figure}[htb]
\includegraphics[scale=0.66]{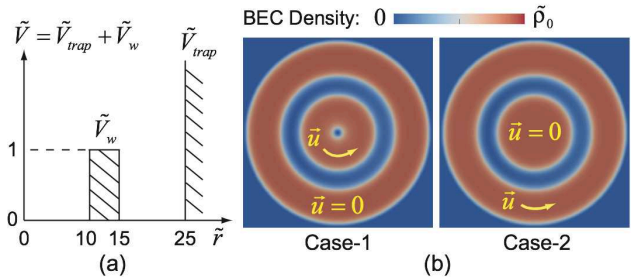}
\caption{(color online). (a) Schematic of the potential $\tilde{V}(\tilde{\textbf{r}},\tilde{t})$ used in the GPE simulation. (b) The initial configurations of the Bose-Einstein condensates.} \label{Fig-1}
\end{figure}

We consider the merging of the BECs at zero temperature whose dynamics can be accurately described by the non-linear Gross-Pitaevskii equation (GPE) \cite{Pitaevskii-2003-book}:
\begin{equation}
i\hbar \frac{\partial \psi}{\partial t}=\left[-\frac{\hbar^2}{2m}\nabla^2+V(\textbf{r},t)+g|\psi|^2\right]\psi,
\label{Eq1}
\end{equation}
where $\psi$ is the condensate wave function, $V$ is the external potential, and $g$ is the coupling constant that measures the strength of the interatomic interactions. This GPE can be reduced to a dimensionless form by rescaling the parameters as $r=\xi{\tilde{r}}$, $t=(\hbar/ng){\tilde{t}}$, and $\psi=(\sqrt{N}/\xi)\tilde{\psi}$, where $\xi=\hbar/\sqrt{2mng}$ is the healing length, $N=\int dS|\psi|^2$ is the total number of particle, and $n=N/S$ is the uniform particle number density:
\begin{equation}
i\frac{\partial \tilde{\psi}}{\partial \tilde{t}}=\left[-\tilde{\nabla}^2+\tilde{V}(\tilde{\textbf{r}},\tilde{t})+\tilde{g}|\tilde{\psi}|^2\right]\tilde{\psi}.
\label{Eq2}
\end{equation}
The dimensionless potential and coupling constant now take the forms $\tilde{V}=V/ng$ and $\tilde{g}=N/(n\xi^2)$. In our simulation, we set $\tilde{V}=\tilde{V}_{trap}+\tilde{V}_w$, where $\tilde{V}_{trap}$ represents the cylindrical hard-wall box potential of the trap and $\tilde{V}_w$ denotes the potential barrier that separates the disc and the ring condensates as shown in Fig.~\ref{Fig-1} (a). $\tilde{V}_w$ has a magnitude of 1. Further increasing $\tilde{V}_w$ does not affect the physics that we will discuss.

Two specific initial configurations of the condensates are considered in our numerical study. In Case-1, the inner disc condensate contains a single vortex point at its center while the outer ring condensate is static. In Case-2, the outer ring condensate carries a supercurrent induced by a virtual vortex point at the center while the inner disc condensate is static. By evolving Eq.~\ref{Eq2} in imaginary time \cite{Chiofalo-2000-PRE}, the steady initial condensate profiles are achieved and are shown in Fig.~\ref{Fig-1} (b). At time $\tilde{t}=0$, we then suddenly remove the energy barrier $\tilde{V}_w$ and let the two condensates merge. These simple yet intriguing condensate configurations are chosen for two reasons: 1) the merging process must involve nontrivial mechanism for angular momentum transfer. Note that the total angular momentum is conserved. Therefore, the flow in the merged condensate cannot be simply induced by the vortex point at the center. How the initially static condensate starts to rotate is, to our knowledge, not reported in literature. 2) These condensate configurations can be easily realized in BEC experiment. For instance, Corman \emph{et al.} \cite{Corman-2014-PRL} and Eckel \emph{et al.} \cite{Eckel-2014-PRX} have utilized the interference patterns of a ring condensate and a disc condensate during free expansion to study Kibble-Zurek mechanism and superfluid weak links. Their setup can be easily adapted to examine the results of our simulation.

\begin{figure}[htb]
\includegraphics[scale=0.8]{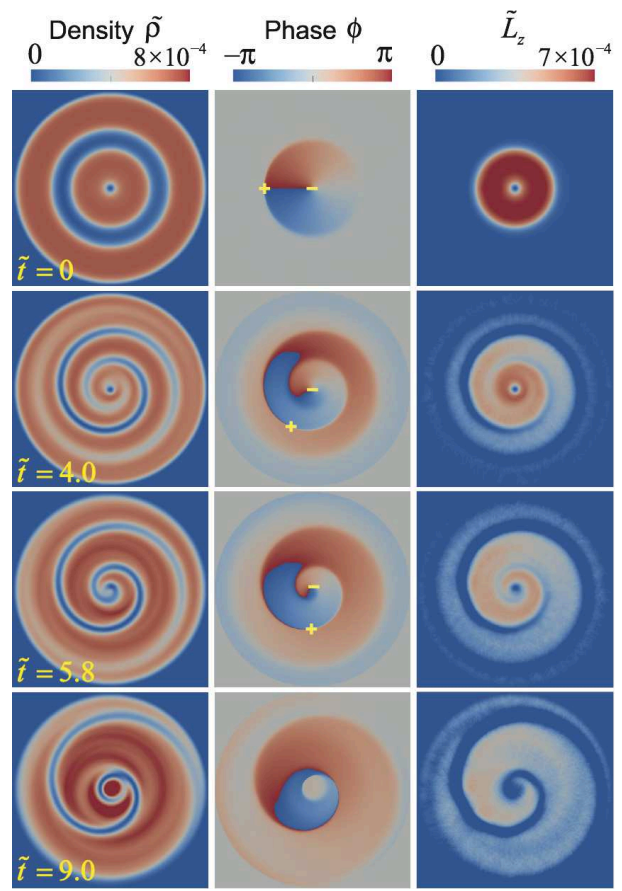}
\caption{(color online). Case-1 representative snapshots showing the time evolution of the BEC density, phase, and angular momentum density $\tilde{L}_z$. The ``+'' and ``-'' signs indicate the locations of the positive and negative vortex points.} \label{Fig-2}
\end{figure}
\begin{figure}[htb]
\includegraphics[scale=0.8]{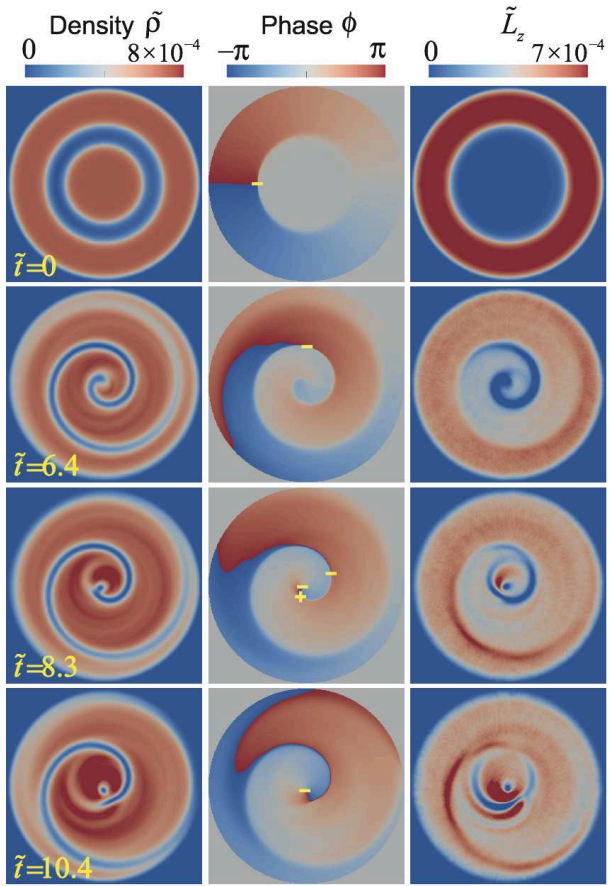}
\caption{(color online). Case-2 representative snapshots showing the time evolution of the BEC density, phase, and angular momentum density $\tilde{L}_z$. The ``+'' and ``-'' signs indicate the locations of the positive and negative vortex points.} \label{Fig-3}
\end{figure}

We carried out our simulation in a regime $\tilde{\textbf{r}}\in[-25,25]\times[-25,25]$ with a mesh grid of $500\times500$ nodes to ensure spatial convergence. The time step $\delta_t$ is chosen to be $1.0\times10^{-4}$. The evolution of the condensate wavefunction during merging is obtained by numerically integrating the Eq.~\ref{Eq2} using an alternating direction implicit method \cite{Press-1992-book}. Representative snapshots of the dynamical evolution of the dimensionless condensate density $\tilde{\rho}=|\tilde{\psi}|^2$ and phase $\phi$ in Case-1 and Case-2, following the removal of the potential barrier $\tilde{V}_w$, are shown in Fig.~\ref{Fig-2} and Fig.~\ref{Fig-3}, respectively. In both cases, we observe the emergence of a spiral stripe with depleted condensate density and with an abrupt phase step $\triangle\phi$ across the stripe boundary. This stripe is indeed a dark soliton, similar in nature to those ring dark solitons identified in the expansion of disc and annular condensates in two dimensional space \cite{Yang-2007-PRA, Yang-2008-PRA, Toikka-2014-JPB}. The boundary of the soliton moves at a speed that is determined by the phase step. For $\triangle\phi=\pi$, the soliton has zero velocity, zero density at its center, and has a width on the order of $\xi$. When $\triangle\phi$ decreases, the soliton becomes shallower and wider, and its speed increases \cite{Jackson-1998-PRA}. The unique spiral shape of the dark soliton seen in our simulation is due to the relative phases between the initial disc and ring condensates. For instance, in Case-1, the ring condensate has a uniform phase while the inner disc has a phase winding of $2\pi$. There is a point across which the relative phase between the disc and the ring condensates changes sign. The soliton then develops two ends with one end spirals in and the other extends out. The different chirality of the spiral solitons in Case-1 and Case-2 indeed reflects the different relative phase winding between the two condensates.


Interesting interactions between the spiral dark soliton and quantized vortex points are also observed. We first note that besides the physical vortices, the terminal point of the phase branch-cut line is deemed as a phase vortex in GPE simulation which does not induce rotational flow but has a phase winding of $2\pi$ around it. In both cases, the dark soliton stripe quickly develops a sharp inner end that spirals towards the center of the condensate. In Case-1, at about $\tilde{t}=5.8$, the negative vortex point initially located at the center merges into the dark soliton. Subsequently, this vortex point moves along the soliton stripe and annihilates with the positive phase vortex, rendering the condensate completely vortex free, as depicted in Fig.~\ref{Fig-2} at $\tilde{t}=9.0$. In Case-2, as the sharp inner end spirals in, the local curvature radius of the soliton stripe becomes comparable to $\xi$. Snake instability then occurs \cite{Mamaev-1996-PRL, Theocharis-2003-PRL, Ma-2010-PRA} and a pair of positive and negative vortex points are nucleated at $\tilde{t}=7.2$. The negative vortex point peels off from the soliton while the positive one sits inside the stripe (see Fig.~\ref{Fig-3} at $\tilde{t}=8.3$). This positive vortex point then moves along the soliton stripe and annihilates with the phase vortex, rendering the condensate with one single negative vortex point at the center. In long time evolution, the solitons in both cases eventually decay into vortices via snake instability (see the movies in the Supplementary Materials).

\begin{figure}[htb]
\includegraphics[scale=0.75]{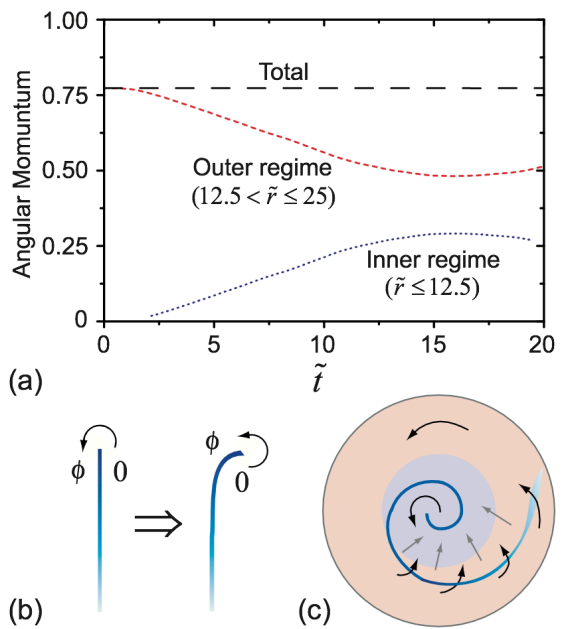}
\caption{(color online). (a) The time evolution of the integrated angular momentum for Case-2. (b) Schematic diagrams showing how the sharp endpoint of a dark soliton stripe can induce rotational motion in a condensate and how such motion bends the soliton stripe. c) Schematics illustrating the mass and angular momentum transfer in Case-2.} \label{Fig-4}
\end{figure}

We have also studied the time evolution of the dimensionless angular momentum density $\tilde{L}_z$, defined as
\begin{equation}
\tilde{L}_z=\frac{\xi^2}{\hbar N}\left(\psi^*\hat{L}_z\psi\right)=\frac{1}{i}\tilde{\psi}^*(\tilde{x}\frac{\partial}{\partial \tilde{y}}-\tilde{y}\frac{\partial}{\partial \tilde{x}})\tilde{\psi}.
\end{equation}
As shown in Fig.~\ref{Fig-2} and Fig.~\ref{Fig-3}, the angular momentum initially is confined to the rotating condensate. During the merging, the angular momentum spreads to the initially static condensate regime along the spiral channel formed by the soliton stripe. The time evolution of the integrated dimensionless angular momentum, for Case-2 as an example, is shown in Fig.~\ref{Fig-4} (a). We see clearly that the angular momentum is transferred from the outer ring regime ($12.5<\tilde{r}\leq25$) to the inner disc regime ($\tilde{r}\leq12.5$) while the total angular momentum is conserved. One may think that this transfer occurs naturally as the condensate flows from the rotating regime through the spiral channel to the initially static regime. However, this is not true. In Case-2, the flow in the initially rotating condensate is counterclockwise which cannot enter the outward spiral channel formed by the soliton. The situation is similar for Case-1. Therefore, the rotation in the initially static condensate must be induced by a different mechanism that should be effective even in the absence of vortices.

Observing the abrupt phase step $\triangle\phi$ across the dark soliton boundary, we realize that for a soliton stripe with a sharp endpoint, there must also be a phase winding of $\triangle\phi$ around this endpoint, as illustrated in Fig.~\ref{Fig-4} (b). Such a phase winding actually leads to a rotational motion in the condensate, making the sharp endpoint effectively a ``vortex point'' that carries a fraction of a quantized circulation given by $(\frac{\triangle\phi}{2\pi})\kappa$. Note that the actual circulation around the endpoint is still zero due to the opposite phase velocity inside the density depleted soliton regime. When the soliton is nearly black (i.e., $\triangle\phi\rightarrow\pi$), mass flow through the soliton boundary is prohibited. In this case, the flow induced by the sharp endpoint transports condensate mass from one side of the soliton stripe to the other side, which leads to spontaneous curving of the sharp end of the soliton stripe. The mechanism for angular momentum transfer observed in our simulation can be identified now. Let us again consider Case-2. As shown in Fig.~\ref{Fig-4} (c), the inner sharp end of the spiral soliton moves toward the center and induces a counterclockwise flow in the inner regime, allowing the inner condensate to rotate. This induced flow carries the condensate mass from the inner regime to the outer regime through the spiral channel. At the meanwhile, this outward flow leads to a phase increment along the soliton boundary, which consequently causes a phase gradient along the radial direction that drives an inward mass flow (see the phase plot at $\tilde{t}=6.4$ in Fig.~\ref{Fig-3}). In the shallow tail part of the soliton stripe, the condensate density in the soliton is not depleted and mass flow from the outer regime through the soliton boundary towards the inner regime becomes significant. As a consequence, a mass circulation between the inner and the outer regimes is formed in the condensate, which effectively mixes the condensate and transports angular momentum between these two regimes.

\begin{figure}[htb]
\includegraphics[scale=0.7]{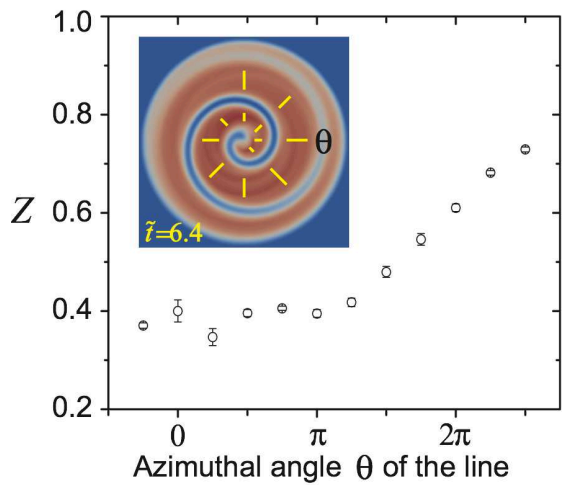}
\caption{(color online). Variation of the calculated vortex charge number $Z$ in the condensate along the solid yellow lines that are shown in the inset.} \label{Fig-5}
\end{figure}

To support our model, we have examined quantitatively the flow field around the spiral dark soliton. Let us take the snapshot in Case-2 at $\tilde{t}=6.4$ as an example, since in this case the flow induced by the sharp inner endpoint of the soliton is not affected by any nearby physical vortices and there is a clear phase winding around the endpoint. To focus on the rotational motion, we introduce a vortex charge parameter $Z$, defined as $Z=\frac{m}{\hbar}|\textbf{r}\times\textbf{v}(\textbf{r})|=\frac{m}{\hbar}rv_{\theta}$, where $v_{\theta}$ is the velocity along the azimuthal angle direction. For a flow field induced by a vortex point at the origin, $Z$ is constant everywhere and equals to the winding number of the vortex. We have calculated the $Z$ values in the condensate along the solid lines as shown in Fig.~\ref{Fig-5}. The error bars represent the variations of $Z$ along these solid lines. The small variation of $Z$ suggests that $v_{\theta}$ scales as $1/r$ along these lines. We see that near the sharp inner endpoint, the $Z$ values are about 0.4 which indeed agrees well with the measured phase step across the soliton boundary towards the inner end (i.e., $\triangle\phi\simeq0.8\pi$). The flow induced by the endpoint in this regime is protected since the nearby soliton boundary has fairly depleted density such that the mass flow through the soliton boundary is weak. In areas where there is appreciable mass flow across the soliton boundary from the outer regime, $Z$ starts to increase and approaches one, a value that is expected for the flow in the initial outer ring condensate.

We have also studied the merging of the condensates with the outer ring carrying a supercurrent with a winding number greater than one. Multiple spiral soliton stripes are seen to emerge in this case, and the number of the spiral solitons matches the winding number. The underlying mechanism for angular momentum transfer is still similar to what we have discussed. More details will be reported in a later publication. 

In summary, our study has revealed that the sharp endpoints of soliton stripes in 2D condensates are nontopological phase defects that can induce rotational motion in the condensates like vortex points, but with a fraction of a quantized circulation that matches the phase step across the soliton boundary. This effect should also exist in 3D condensates. For instance, a disc-shaped soliton with a sharp edge in 3D BECs should have a phase winding around its edge, inducing rotational motion effectively like a vortex ring. Indeed, Shomroni \emph{et al.} have reported the creation of disc dark solitons in 3D cigar-shaped condensates ~\cite{Shomroni-2009-NP}. But in those experiments, the soliton edge is on the BEC surface, which renders no visible rotational motion around the disc edge. Nevertheless, besides facilitating angular momentum transfer in the merging of condensates, endpoints and edges of dark solitons as nontopological phase defects may play important roles in many other quantum transport processes and even in quantum turbulence ~\cite{Vinen-2002-JLTP} where rotational motions are essential.

\begin{acknowledgments}
W. G. acknowledges the support from the National Science Foundation under Grant No. DMR-1507386. W. G. would also like to thank the support from The Japan Society for the Promotion of Science (JSPS) through the Invitational Fellowship Program (No. S17018) and the Osaka City University for hosting his visit. M. T. would like to acknowledge the support by JSPS KAKENHI under Grant No. JP17K05548 and JP16H00807.
\end{acknowledgments}

\end{document}